\documentclass[useAMS,usenatbib]{mn2e}

\usepackage{aas_macros}
\usepackage{hhline}
\usepackage{graphicx}
\usepackage{bmpsize}
\usepackage{amssymb,amsmath}
\usepackage{comment}

\graphicspath{.{/mnt/data/Work/A3376/paper/figs/}}
\def\mjyb{mJy beam$^{-1}$ }
\def\mjyam{mJy arcmin$^{-2}$ }

\def\RRI{$^{1}$}
\def\Victoria{$^{2}$}
\def\Curtin{$^{3}$}
\def\CAASTRO{$^{4}$}
\def\UWA{$^{5}$}
\def\CASS{$^{6}$}
\def\USydney{$^{7}$}
\def\UMelbourne{$^{8}$}
\def\ASTRON{$^{9}$}
\def\SKASA{$^{10}$}
\def\Rhodes{$^{11}$}
\def\CfA{$^{12}$}
\def\ASU{$^{13}$}
\def\ANU{$^{14}$}
\def\Haystack{$^{15}$}
\def\MIT{$^{16}$}
\def\UW{$^{17}$}
\def\UWisc{$^{18}$}
\def\UMichigan{$^{19}$}
\def\Tata{$^{20}$}
\def\NRAO{$^{21}$}

\title[MWA Observations of A3376]{An analysis of the halo and relic radio emission from Abell 3376 from Murchison Widefield Array observations}
\author[George et. al.]{
L.~T.~George\RRI\thanks{lijo@rri.res.in}, 
K.~S.~Dwarakanath\RRI, 
M.~Johnston-Hollitt\Victoria, 
N.~Hurley-Walker\Curtin, 
\newauthor{
L.~Hindson\Victoria, 
A.~D.~Kapi\'nska\CAASTRO$^,$\UWA, 
S.~J.~Tingay\Curtin$^,$\CAASTRO, 
M.~Bell\CASS, 
J.~R.~Callingham\CAASTRO$^,$\CASS$^,$\USydney, 
}
\newauthor{
Bi-Qing For\UWA, 
P.~J.~Hancock\Curtin$^,$\CAASTRO, 
E.~Lenc\USydney, 
B.~McKinley\UMelbourne, 
J.~Morgan\Curtin, 
A.~Offringa\ASTRON, 
}
\newauthor{
P.~Procopio\UMelbourne, 
L.~Staveley-Smith\UWA, 
R.~B.~Wayth\Curtin, 
Chen~Wu\UWA, 
Q.~Zheng\Victoria, 
}
\newauthor{
G.~Bernardi\SKASA$^,$\Rhodes$^,$\CfA, 
J.~D.~Bowman\ASU, 
F.~Briggs\ANU, 
R.~J.~Cappallo\Haystack, 
B.~E.~Corey\Haystack, 
}
\newauthor{
A.~A.~Deshpande\RRI,
D.~Emrich\Curtin,
R.~Goeke\MIT,
L.~J.~Greenhill\CfA, 
B.~J.~Hazelton\UW, 
}
\newauthor{
D.~L.~Kaplan\UWisc, 
J.~C.~Kasper\CfA$^,$\UMichigan, 
E.~Kratzenberg\Haystack, 
C.~J.~Lonsdale\Haystack, 
M.~J.~Lynch\Curtin, 
}
\newauthor{
S.~R.~McWhirter\Haystack, 
D.~A.~Mitchell\CAASTRO$^,$\CASS,
M.~F.~Morales\UW, 
E.~Morgan\MIT, 
D.~Oberoi\Tata, 
}
\newauthor{
S.~M.~Ord\Curtin$^,$\CAASTRO, 
T.~Prabu\RRI,
A.~E.~E.~Rogers\Haystack, 
A.~Roshi\NRAO, 
N.~Udaya~Shankar\RRI, 
}
\newauthor{
K.~S.~Srivani\RRI, 
R.~Subrahmanyan\RRI$^,$\CAASTRO, 
M.~Waterson\Curtin$^,$\ANU,
R.~L.~Webster\CAASTRO$^,$\UMelbourne,
}
\newauthor{
A.~R.~Whitney\Haystack, 
A.~Williams\Curtin, 
C.~L.~Williams\MIT 
}
\\
$^{1}$Raman Research Institute, Bangalore 560080, India\\
$^{2}$School of Chemical \& Physical Sciences, Victoria University of Wellington, Wellington 6140, New Zealand\\
$^{3}$International Centre for Radio Astronomy Research, Curtin University, Bentley, WA 6102, Australia\\
$^{4}$ARC Centre of Excellence for All-sky Astrophysics (CAASTRO)\\
$^{5}$International Centre for Radio Astronomy Research, University of Western Australia, Crawley, WA 6009, Australia\\
$^{6}$CSIRO Astronomy and Space Science (CASS), PO Box 76, Epping, NSW 1710, Australia\\
$^{7}$Sydney Institute for Astronomy, School of Physics, The University of Sydney, NSW 2006, Australia\\
$^{8}$School of Physics, The University of Melbourne, Parkville, VIC 3010, Australia\\
$^{9}$ASTRON, 7990 AA Dwingeloo, The Netherlands\\
$^{10}$Square Kilometre Array South Africa (SKA SA), 3rd Floor, The Park, Park Road, Pinelands, 7405, South Africa\\
$^{11}$Department of Physics and Electronics, Rhodes University, PO Box 94, Grahamstown, 6140, South Africa\\
$^{12}$Harvard-Smithsonian Center for Astrophysics, 60 Garden Street, Cambridge, MA 02138, USA\\
$^{13}$School of Earth and Space Exploration, Arizona State University, Tempe, AZ 85287, USA\\
$^{14}$Research School of Astronomy and Astrophysics, Australian National University, Canberra, ACT 2611, Australia\\
$^{15}$MIT Haystack Observatory, Westford, MA 01886, USA\\
$^{16}$Kavli Institute for Astrophysics and Space Research, Massachusetts Institute of Technology, Cambridge, MA 02139, USA\\
$^{17}$Department of Physics, University of Washington, Seattle, WA 98195, USA\\
$^{18}$Department of Physics, University of Wisconsin--Milwaukee, Milwaukee, WI 53201, USA\\
$^{19}$Department of Atmospheric, Oceanic and Space Sciences, University of Michigan, Ann Arbor, MI 48109, USA\\
$^{20}$National Centre for Radio Astrophysics, Tata Institute for Fundamental Research, Pune 411007, India\\
$^{21}$National Radio Astronomy Observatory, Charlottesville and Greenbank, USA\\
}

\pagerange{\pageref{firstpage}--\pageref{lastpage}} \pubyear{2014}

\begin{document}
	
	\label{firstpage}

	\maketitle
	\clearpage
    \begin{abstract}
    We have carried out multiwavelength observations of the near-by ($z=0.046$) rich, merging galaxy cluster
    Abell 3376 with the Murchison Widefield Array (MWA). 
    As a part of the GaLactic and Extragalactic All-sky MWA survey (GLEAM), this cluster was observed at 88, 118, 154, 188 and 215 MHz. 
    The known radio relics, towards the eastern and western peripheries of the cluster, were detected at all the frequencies. 
    The relics, with a linear extent of $\sim$ 1 Mpc each, are separated by $\sim$ 2 Mpc.
    Combining the current observations with those in the literature, we have obtained the spectra of these relics over the frequency range 80 -- 1400 MHz. 
    The spectra follow power laws, with $\alpha$ = $-1.17\pm0.06$ and $-1.37\pm0.08$ for the west and east relics, respectively ($S \propto \nu^{\alpha}$).
    Assuming the break frequency to be near the lower end of the spectrum  we estimate the age of the relics to be $\sim$ 0.4 Gyr. 
    No diffuse radio emission from the central regions of the cluster (halo) was detected.
    The upper limit on the radio power of any possible halo that might be present in the cluster is a factor of 35 lower than that expected from the radio power and X-ray luminosity correlation for cluster halos.
    From this we conclude that the cluster halo is very extended ($>$ 500 kpc) and/or most of the radio emission from the halo has decayed. 
    The current limit on the halo radio power is a factor of ten lower than the existing upper limits with possible implications for models of halo formation.
    \end{abstract}


    \section{Introduction}
    Galaxy clusters are some of the largest known gravitationally bound structures in the Universe composed primarily of dark matter along with galaxies and the X-ray emitting Intra-Cluster Medium (ICM) gas. 
    Their masses are $\sim 10^{15}M_{\odot}$ while their sizes are $\sim$ Mpc. 
    According to the theory of hierarchical structure formation, galaxy clusters are formed by mergers of several smaller sub-clusters and groups driven by gravity. 
    The process of cluster evolution through mergers is still ongoing and is one of the most energetic phenomena in the Universe with energies 10$^{63}-10^{64}$ erg \citep{hoeft08}.

    In some merging clusters Mpc-scale diffuse synchrotron radio emission, in the form of {\it radio halos} (in the central regions of the cluster) and/or {\it radio relics} (towards the peripheries), have been detected. 
    The radio halos have a typical surface brightness of a few \mjyam at 1.4 GHz (see \citet{feretti12} and \citet{brujones14} for recent reviews). 
    Radio halos usually have a smooth morphology while relics are found to be elongated or arc-like in nature. 
    The fraction of galaxy clusters which host such structures is rather small. 
    Of all the known galaxy clusters below a redshift of 0.2, $\sim$ 5\% contain radio halos and relics \citep{giovannini99}. 
    This fraction increases to $\sim$ 35\% for higher X-ray luminosity clusters ($L_X \gtrsim 10^{45}$ erg s$^{-1}$) \citep{venturi08} which have correspondingly higher radio power as well \citep{liang2000,bacchi03,cassano06,cassano07,brunetti07,brunetti09,rudnick09}. 
    In a more recent study of galaxy clusters with redshift $0.2<z<0.4$, \citet{kale13} estimated that 23\% of galaxy clusters with X-ray luminosity $L_X > 5\times10^{44}$ erg s$^{-1}$ host radio halos. 
    For clusters with $L_X > 8\times10^{44}$ erg s$^{-1}$ the fraction increases to 31\%. 
    It has been observed, however, that highly disturbed, merging clusters are much more likely to host halos and relics than relaxed clusters \citep{buote01,cassano10b}. 
    This led to the scenario that major cluster mergers could provide sufficient energy to the electrons in the ICM which, in the presence of existing magnetic field, produce synchrotron radiation that is detected as halos and relics. 
    It should be mentioned here that even though both halos and relics are found predominantly in disturbed clusters, the processes responsible for their generation are thought to be different, as explained below.

    One of the currently accepted models for the generation of radio halos is {\it turbulent acceleration} \citep{petrosian01,brunetti01}. 
    During the process of a cluster merger, turbulence in the ICM gas re-accelerates the electron population of the ICM to ultra-relativistic energies ($\gamma\gtrsim$ 1000, where $\gamma$ is the Lorentz factor). 
    This turbulence also amplifies the magnetic field to a few $\mu$G \citep{carilli02,subra06}. 
    The synchrotron radiation from these electrons could be seen in the form of radio halos from the cluster centre.

An alternative to the turbulent acceleration model is the \textit{hadronic} (secondary) model \citep{dennison80, blacola99, pfrosslin04}, which posits that collisions between relativistic and thermal protons in the ICM are responsible for the generation of relativistic electrons. Proton-proton collisions result in the formation of pions. Neutral pions decay to produce gamma rays whereas charged pions decay to produce electrons and positrons. Since relativistic protons in galaxy clusters have a lifetime greater than that of the cluster itself, there will always be a fraction of the relativistic electrons in the ICM produced in such a manner. However, based on radio observations of massive, X-ray luminous galaxy clusters, \citet{brunetti07} claimed that relativistic protons contribute less than a few percent of the energy in particles in galaxy clusters. Recently, observations of galaxy clusters at GeV energies were carried out using the \textit{Fermi} and Cherenkov telescopes \citep{aharonian09b,aharonian09a,ackermann10,aleksic10}. No gamma rays were detected and only upper limits to the emission were placed. Therefore, any contribution to synchrotron emission due to relativistic protons (indirectly) would be very small. It is believed that the level of radio emission produced in this process (secondary model) is about a factor of 10 below the radio emission produced in the turbulence model \citep{brulaz11}.

Radio relics, on the other hand, are created as a consequence of shock fronts travelling outwards from the cluster centre after a merger. The acceleration of electrons in the periphery of the cluster is best explained by the diffusive shock acceleration (DSA) theory \citep{blandford87,jones91}. According to the theory, shock acceleration is essentially a first-order Fermi acceleration process in which thermal particles are accelerated diffusively. They scatter elastically across the two fronts of a shock repeatedly, gaining energy at each crossing of a front until they get accelerated to very high energies \citep{feretti12}. The best examples for the validity of this theory come in the form of a special class of relics, known as double relics, where two arc-like structures are seen on the opposite sides of the cluster centre. However there are very few of these double relics (e.g. Abell 3667 \citep{mjh03}, Abell 3376 \citep{bagchi06}, CIZA J2242.8+5301 \citep{vanweeren11} etc.) known. Rarer still are clusters that host both halos and double relics, even though both are thought to be a consequence of the same merger process  (e.g. RXC J1314.5-2515 \citep{feretti05}, CIZA J2242.8+5301, El Gordo \citep{lindner14} etc.). The reason for the absence of halos in merging systems is not well understood. One possibility is that the halo was created at the time of the merger but faded away over time. 

Radio halos and relics are systems where different physical processes come into the picture. Detection or non-detection of diffuse radio emission depends on many factors such as the mass ratio of the merging clusters, impact parameter of collision between the clusters, time since the merger, turbulence decay time scales, and synchrotron lifetime of the radiating electrons \citep{brujones14}. In the case of radio relics, the inhomogeneity of the ICM also plays an important part. The dearth of detections of radio halos and radio relics can be explained in many ways. Clusters with lower X-ray luminosity have correspondingly lower radio power and may not be bright enough to be detected by existing telescopes. 
Additionally, if one assumes the turbulent acceleration model that produces the radio halos, then the diffuse radio emission should display an exponential cut-off at higher radio frequencies ($\gtrsim1.4$GHz). As a result most of the extended, diffuse emission will be fainter at higher frequencies \citep{cassano10}.  
Furthermore, large angular extent diffuse emission is also likely to be resolved out at higher frequencies. This is why it is necessary to observe and image galaxy clusters at low radio frequencies with greater surface brightness sensitivity than what GHz receivers can currently offer. Telescopes such as the MWA \citep{lonsdale09,tingay13,bowman13}, LOFAR (LOw Frequency ARray, \citet{vanhaarlem13}) and the upcoming SKA (Square Kilometre Array) \citep{dewdney13} will be crucial in this endeavour. 

With the above motivations in mind, recently, \citet{hindson14} used commissioning data from the MWA to observe the galaxy cluster Abell 3667 at low frequencies (120, 149, 180 and 226 MHz). A3667 has long been known to host two radio relics. Its northern most relic has been discussed in a number of papers spanning 45 years \citep{ekers69,schilizzi75,goss82,rottgering97}, whilst the weaker second relic, though visible in the published 843 MHz images in \citet{rottgering97} was overlooked at that time, and not first mentioned in the literature until it was rediscovered in the 1.4 GHz data and the entire cluster re-imaged \citep{mjh02,mjh03} thus making A3667 the first double relic cluster. Imaging between 1.4 and 2.4 GHz has given rise to suggestions of a radio halo \citep{mjh02,mjh03,carretti13}, a possible bridge connecting the two relics \citep{carretti13} or more recently a mini-halo off-set from the cluster centre \citep{riseley15}. Assuming typical spectral characteristics, the putative halo and bridge were expected to be easily visible in the lower frequency MWA imaging. 
However, the MWA imaging did not confirm the detection of either structure. Based on this MWA imaging it has been suggested that A3667 is seen in projection through a large diffuse spur of Galactic emission, which in turn might be responsible for these features \citep{hindson14}. At the sensitivity of the MWA commissioning observations it was not possible to distinguish the Galactic features from a weak halo or bridge that might also be present, though strong limits were placed on the possible emission. In our current observations of the morphologically similar system, A3376, we expect to reach a sensitivity that is a factor of 5 better than the commissioning
observations.

	\subsection{Abell 3376}
	\begin{figure}
		\includegraphics[width=\columnwidth]{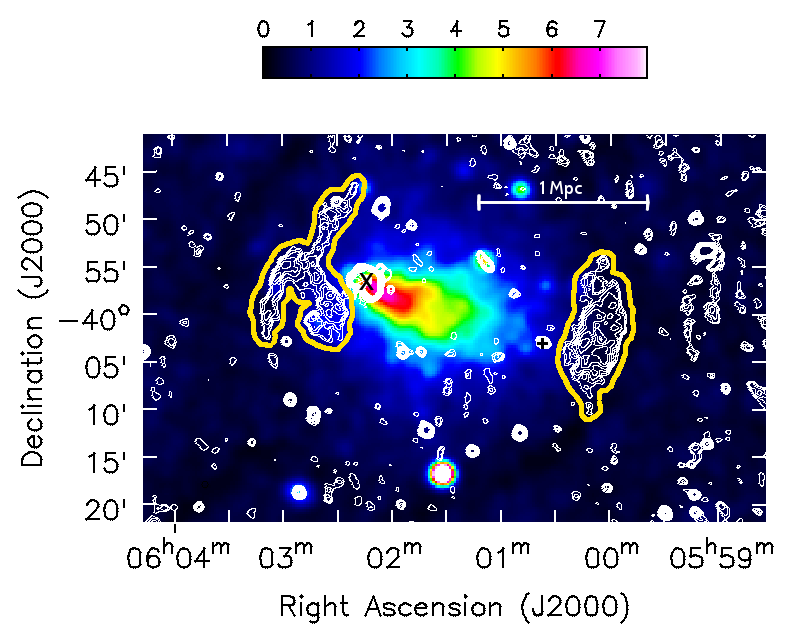}
		\caption{GMRT 325 MHz contours \citep{kale12} on the ROSAT X-ray (0.1-2.4 keV) broadband image. The units of the X-ray image are in counts s$^{-1}$ arcmin$^{-2}$. The contour levels start from 5.8 \mjyb ($3\sigma$) and increase by a factor of $\sqrt{2}$. The resolution of the radio image is $39''\times39''$. The $\times$ and the + represent the BCG2 and BCG1 galaxies, respectively. The yellow solid line (drawn by the authors) marks the boundaries of the east and west relics.}
		\label{g325}
	\end{figure}

	Abell 3376 is a rich, merging galaxy cluster at a redshift of $z = 0.046$ \citep{strood99}. It has an X-ray luminosity of $L_{X [0.1-2.4 \text{ keV}]} = 2.2\times 10^{44}$erg s$^{-1}$ \citep{reiprich02} and a virial mass of $\sim 3.64 \times 10^{14}$ M$_{\sun}$ \citep{girardi98}. Optical observations \citep{escalera94,durret13} indicate the merging status of the cluster by showing the two Brightest Cluster Galaxies (BCG) on the opposite ends of the cluster with the X-ray peak at BCG2 (MRC0600-399) (Fig.~\ref{g325}). The BCG1 is a cD galaxy and is considered to be the centre of the main cluster. The BCG2 is an elliptical and its radio counterpart is a bent-jet galaxy. \citet{machado13} simulated the merger of A3376 and compared their simulation results with the X-ray morphology of the cluster. Based on these comparisons they suggest that the merger in A3376 occurred $\sim$ 0.5 Gyr ago between subclusters having a mass ratio of 6:1. Very Large Array observations of this cluster at 1.4 GHz \citep{bagchi06} showed the presence of two arc-like structures in the eastern and western peripheries of the cluster, $\sim$ 1 Mpc away from the cluster centre. These relics are assumed to be due to synchrotron emission from the ICM as a consequence of shock waves travelling outward from the cluster centre soon after the merger. Further radio observations at lower frequencies (150 and 325 MHz) using the Giant Meterwave Radio Telescope \citep{kale12} detected the relics but not a radio halo from the central regions of the cluster (Fig.\ref{g325}). The east relic is found to be polarised up to 30\% with aligned magnetic field vectors, whereas the west relic is polarised between 5 and 20\% and the field vectors are not aligned in any particular direction \citep{kale12}.
	
	The motivations for the current study are twofold. First, to detect the radio relics at low frequencies and establish their spectra over 80 -- 1400 MHz. Second, to detect or put stringent upper limits on the radio power of a possible halo in the central regions of the cluster, which has eluded detection so far. The paper is organised as follows: In section 2 we discuss observations and data analysis. Section 3 contains the results and in section 4 we discuss the implications of these results. We conclude with the summary in Section 5. In this paper we adopt a $\Lambda$--CDM cosmology with $H_0 = 70$ km s$^{-1}$ Mpc$^{-1}$, $\Omega_{M} = 0.3$ and $\Omega_{\Lambda} = 0.7$.
	
\section{Observations and Analysis}

    \subsection{The GLEAM Survey}
    The observations of A3376 were carried out as part of GaLactic and Extragalactic All-Sky MWA (GLEAM) Survey (Wayth et al., submitted).
    There are five observing frequency ranges: 72.3--103.04 MHz, 103.04--133.76 MHz, 138.88--169.6 MHz, 169.6--200.32 MHz and 200.320--231.04 MHz, each having a bandwidth of 30.72 MHz. 
    The data were recorded with an integration time of 0.5 second and a spectral resolution of 40 kHz. 
    The frequency range 134--138 MHz was avoided due to the presence of satellite RFI in that range.
	
    The GLEAM survey was carried out in the drift-scan mode with the antennas pointing to one of the seven declination settings ($\delta = +18.6^{\circ}, +1.6^{\circ},-13.0^{\circ}, -26.7^{\circ}, -40.0^{\circ}, -55.0^{\circ}, -72^{\circ}$) and letting the sky drift overhead. 
    A single scan consists of observations for 112 seconds at each of the five frequencies available and then cycling through them. 
    The data used for imaging A3376 in this paper were taken during November 6--7, 2013 while covering the RA range 0--8h at declination $\delta = -40^{\circ}$. 
    This declination pointing is within a few arcminutes of the centre of the cluster (Fig.~\ref{g325}).

    \subsection{Data reduction} 
    In principle, one could use the specialised pipeline software that has been set up to analyse the observations from the GLEAM survey (Hurley-Walker, in prep.). 
    However, with the scope to try to analyse the MWA observations using a standard radio astronomy software we imaged the cluster at two frequencies viz. 
    154 and 215 MHz using CASA \footnote{http://casa.nrao.edu} (Common Astronomy Software Applications). 
    We chose these two frequencies because at 215 MHz the highest angular resolution in the GLEAM frequency range is achieved ($151''\times145''$) and at 154 MHz the MWA images obtained can be directly compared with the existing GMRT observations at 150 MHz. 
    The images at 154 and 215 MHz, processed with CASA, are further compared with the pipeline processed images at the same frequencies. 
    For the remaining frequencies (88, 118 and 188 MHz) we use the pipeline processed images.
    
    The raw data from A3376 scans were converted to measurement sets readable by CASA using the COTTER software \citep{offringa15}. 
    This software also performed some initial flagging and RFI removal using the AOFlagger library \citep{offringa10,offringa12}. 
    The source 3C161 was used to create the normalised bandpass tables (task \textsc{bandpass} in CASA)  which were applied to all the A3376 scans.
    In all, a total of seven scans of A3376 were used -- the transit scan and three scans on either side of the source's transit through the meridian -- which covered a total time range of $\pm30$ minutes around transit at each frequency. 
    The scans beyond this time range were not considered due to the falling gain of the primary beam (i.e. the beam response was falling $<$ 50\%). 
    Note that the full widths at half maxima of the primary beams of the MWA tiles are $\sim 25^{\circ}$ at 154 MHz \citep{nhw14}.
    The bandpass corrected measurement sets were then averaged in time and frequency to 8 s and 400 kHz, respectively.
    Each scan was then individually imaged and deconvolved in CASA using the task \textsc{clean}.
    Each image produced (2048$\times$2048 pixels) covered an area of $34.13\times34.13$ deg$^2$ in each axis with a pixel size of 1$'.$
	Deconvolution was performed using the Cotton-Schwab algorithm \citep{schwab84} and widefield effects were taken into account using the $w$-projection algorithm \citep{cornwell08} by setting the number of projection planes in \textsc{clean} to 256.

    A variety of robust weighting schemes were tried to make images of A3376. In order to reach a balance between sensitivity and resolution we used ``Briggs'' weighting \citep{briggs95} with robust value 0 for each scan at both polarisations. Note that Briggs weighting with values $-5$ and $+5$ correspond to purely uniform and purely natural weighting, respectively. The images were then sent through four rounds of phase self-calibration, at the end of which the phase errors had reduced to an rms of $0.2^{\circ}$. No further improvement was seen with subsequent phase and amplitude self-calibration. The XX and YY polarisations were imaged separately for each scan. They were then corrected for primary beam effects \citep{sutinjo14}.

	Following the above procedure a total of seven images were produced for A3376 for each polarisation. These images were examined separately and their rms values were estimated. Furthermore, the positions of unresolved sources in the GMRT 150 MHz image within a degree from A3376 were compared with their respective positions in the seven MWA 154 MHz images. Across these seven images, the positional offsets for all these sources were constant to within 10\% of the MWA synthesized beam size at 154 MHz. At each frequency, the individual primary beam corrected XX and YY images were weighted using the inverse variance of the rms in the respective images and then combined. The final XX and YY images were added together to produce the full total intensity (Stokes I) image.

      	\begin{figure}   	
   		\includegraphics[width=\columnwidth]{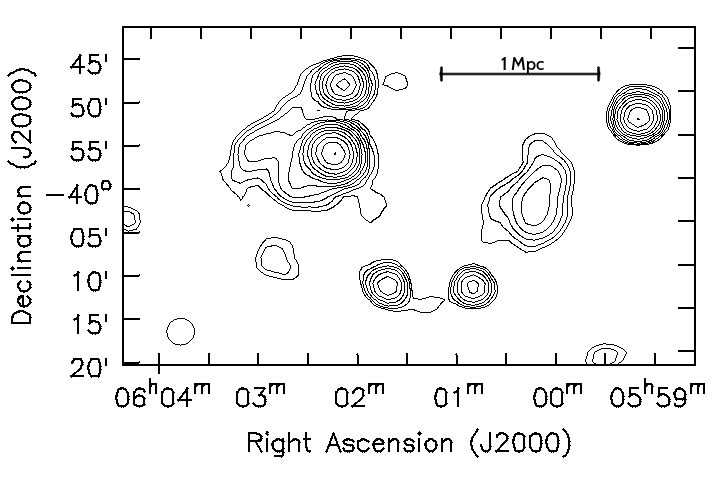}
   		\includegraphics[width=\columnwidth]{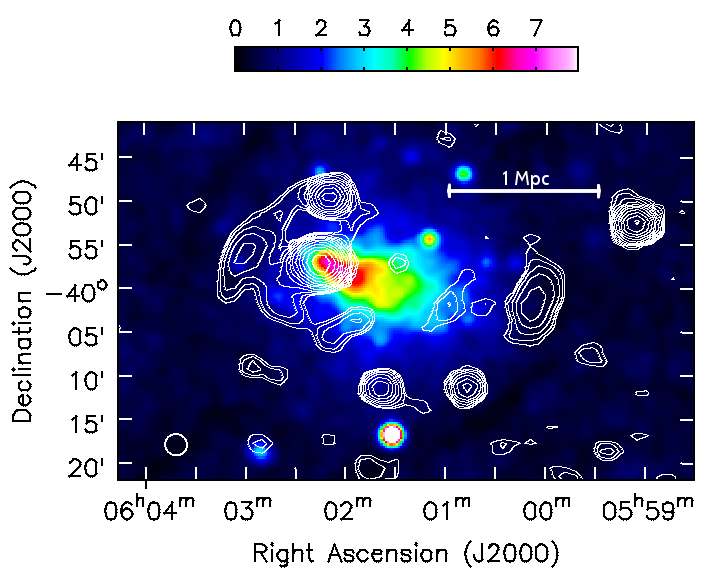}
   		\caption{MWA 154 MHz (top) and 215 MHz (bottom) contour images. The 215 MHz contours are overlaid on ROSAT X-ray (0.1-2.4 keV) image. The units of the X-ray image are in counts s$^{-1}$ arcmin$^{-2}$. Both images have a robust weighting of 0. The 154 MHz image has a resolution of $190''\times184''$ $-85.6^{\circ}$ while the 215 MHz image has a resolution of $151''\times 145'', $ $68.2^{\circ}$. The synthesised beam is indicated at the bottom left corner in both the images. Contours for the 154 MHz and 215 MHz images start at 54 \mjyb ($3\sigma$) and 33 \mjyb ($3\sigma$), respectively, and increase by a factor of $\sqrt{2}$.}
   		\label{fig:m150_m215}
   	\end{figure}

    For the remaining frequencies (88, 118 and 188 MHz), we utilised the pipeline-processed GLEAM images, described in detail in Hurley-Walker et. al (in prep). For each scan, the basic steps of this pipeline are as follows,
\begin{itemize}
\item{Perform a first-pass calibration using a single bright source, applied to all observations \citep{nhw14};}
\item{Using \textsc{WSClean} \citep{offringa14}, invert and clean multi-frequency-synthesis images in instrumental Stokes, across the full bandwidth, stopping at the first negative clean component;}
\item{Use the MWA primary beam model \citep{sutinjo14} to transform to astronomical Stokes;}
\item{Set Q, U and V to zero (i.e. assume that the sky is unpolarised) and use the same beam model to transform back to instrumental Stokes;}
\item{Use the sky model in a self-calibration loop to update the gain solutions;}
\item{Form new multi-frequency synthesis images.}
\end{itemize}

These individual images were then combined to produce the final total intensity images exactly as described earlier.

    \begin{table*}
		\centering 
		\begin{tabular}{cccccc}
		\hline
		Object		&	\multicolumn{5}{c}{Flux Density (mJy)}			\\
					&	88	MHz		&	118	MHz		&	154	MHz		&	188	MHz		&	215	MHz			\\
		\hline
		East Relic	&	$5476\pm675$&	$3886\pm577$& 	$2508\pm465$&	$1508\pm169$& 	$2052\pm350$	\\
		West Relic	&	$2805\pm348$&	$1926\pm288$&	$1351\pm253$&	$1172\pm134$&	$609\pm108$	\\
		Central Region&	$48.5\pm47$ &	$59.5\pm28.2$	&$22\pm18$	&	$20\pm12$	&	$17\pm11$				\\
		\end{tabular}
		\caption{The integrated flux densities for the east and west relics along with their uncertainties are given in the first two rows. The third row gives the mean value of the flux density from the central region at different frequencies. These mean values represent upper limits to halo emission in the MWA synthesised beam ($\sim$160 kpc at 154 MHz). The errors quoted on these mean values for the central region are the rms values at the corresponding frequencies.}
		\label{table:flux}
	\end{table*}

	\subsection{Calibration}
   The absolute flux density calibration of the MWA images was tied to those of the GMRT images at 150 and 325 MHz \citep{kale12}. 
For GMRT calibration, 3C147 and 3C48 as primary calibrators. The Perley-Butler (2010) scale was used for absolute flux density calibration. The secondary calibrators used in the GMRT 150 and 325 MHz observations were 0521-207 and 0837-198, respectively. The flux densities of these two calibrators obtained from GMRT at the two frequencies agree to within 10\% of their expected values based on the VLA calibrator list \citep{kale12}.
   
   The MWA 154 MHz image was calibrated using the GMRT 150 MHz image. Unresolved sources in the GMRT 150 MHz image within a degree from A3376 were used to estimate an average scaling factor between the GMRT 150 MHz and the MWA 154 MHz images 
   
   To calibrate the MWA 215 MHz image, the spectral indices of the same unresolved sources were estimated between the GMRT 150 MHz and 325 MHz images. The extrapolated flux densities at 215 MHz were estimated for these sources. The MWA 215 MHz image was then scaled according to the average value of the
scaling factor. The images at 88, 118, and 188 MHz were also calibrated in a manner similar to the one used for calibrating the MWA 215 MHz image. 

	The images at 154 and 215 MHz made from CASA and from pipeline-processing were compared. At 154 MHz, the rms of the CASA processed image was 18 \mjyb (Table~\ref{table:flux}) while that of the pipeline processed image was 13 \mjyb. At 215 MHz, the rms of the CASA image was 11 \mjyb (Table~\ref{table:flux}) while the pipeline image had an rms of 10 \mjyb. The rms of the pipeline images could be expected to reduce further by mosaicking in narrower frequency channels, resulting in fewer excluded images and thus a higher signal-to-noise ratio (Hurley-Walker et al., in prep).

   	\begin{figure*}
   		\begin{minipage}{\columnwidth}
			\includegraphics[width=0.7\columnwidth,angle=270]{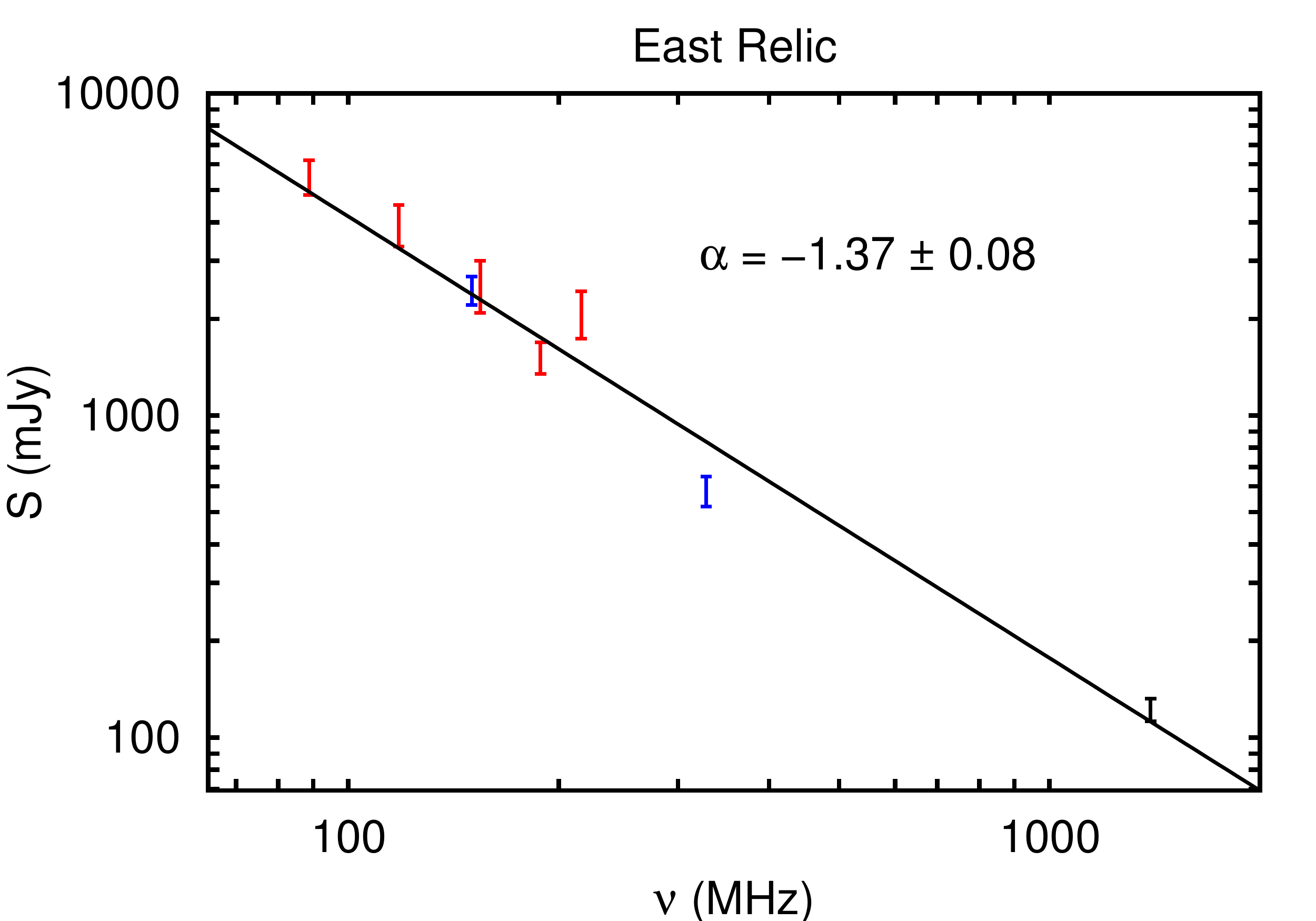}
   		\end{minipage}
   		\begin{minipage}{\columnwidth}
   			\includegraphics[width=0.7\columnwidth,angle=270]{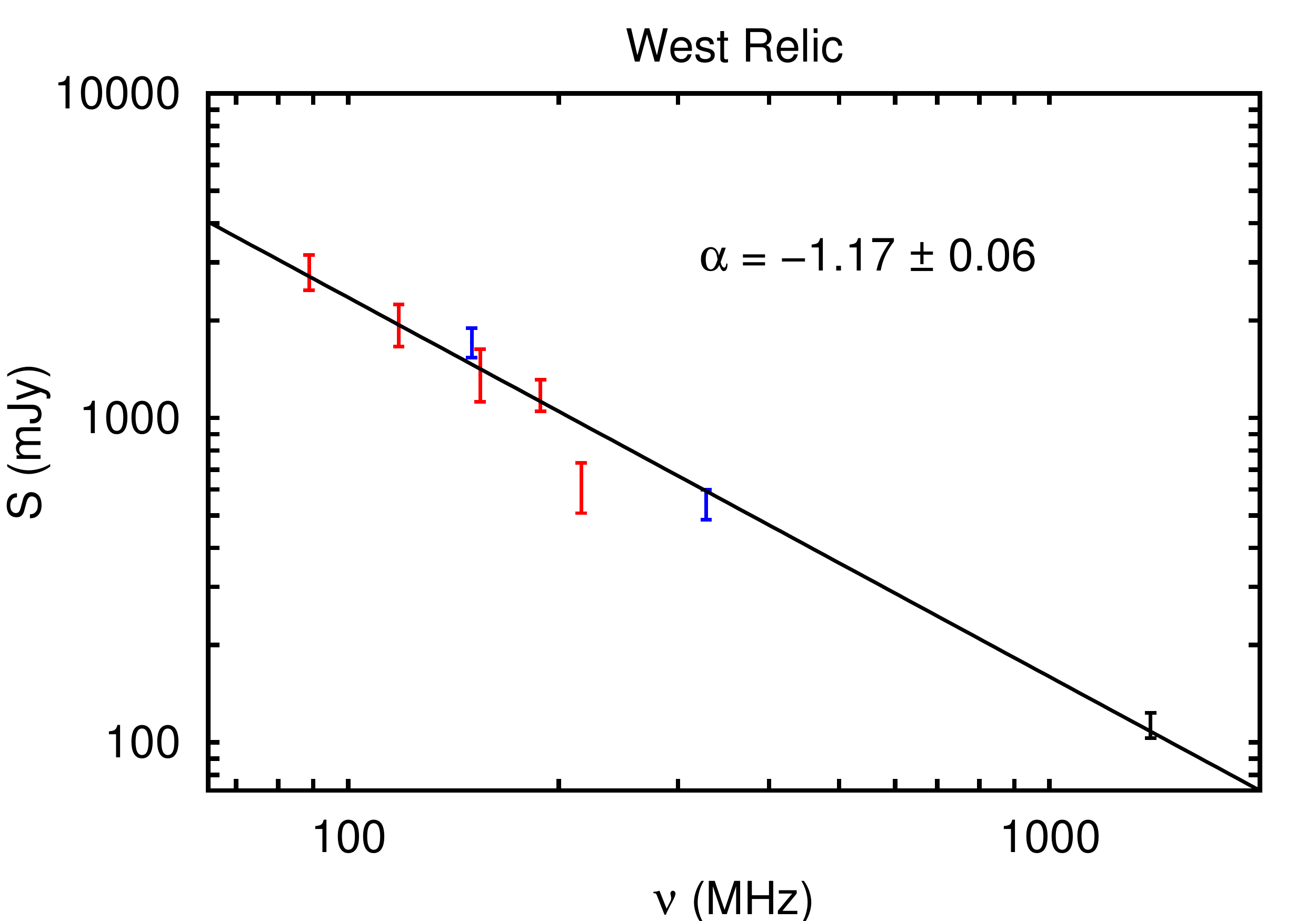}
   		\end{minipage}
   		\caption{Spectra for the east and west relics. These spectra include measurements from the GMRT observations at 150 and 325 MHz \citep{kale12} (blue), from the VLA observation at 1.4 GHz \citep{bagchi06} (black) and from the current MWA observations at 88, 118, 154, 188, 215 MHz (red). Also displayed are the best fit lines for the respective data points as well as the spectral index estimates.}
	  	\label{fig:spec_rel}
   	\end{figure*}
   	
\section{Results}
The MWA images of the Abell 3376 region at 154 and 215 MHz are shown in Fig.~\ref{fig:m150_m215}. 
The images at the other MWA frequencies are not presented here as they are essentially convolved versions of the above images.
The Abell 3376 images at 154 and 215 MHz show two prominent relics just outside the region of X-ray emission: one toward the east and the other toward the west of the X-ray emission (compare Fig. \ref{g325} and \ref{fig:m150_m215}). 
The two relics are separated by $\sim$ 2 Mpc ($\sim30'$) from each other. 
Both the east and west relics were detected at all the MWA frequencies. 
No diffuse radio emission in the central regions of the cluster (halo) was detected down to the GLEAM sensitivity. 
Table \ref{table:flux} lists the integrated flux densities of the relics and the upper limits on the flux densities of a possible halo in the region between the relics.

There are two primary sources of errors in the estimation of the flux densities of the relics in the MWA images. First, there is an error due to the uncertainties in the flux densities of the unresolved sources used for calibration of the MWA images. This error is estimated to be $\sim$10\%. Second, since the relics are extended sources in the MWA observations, the errors in their flux density estimations will be the rms in the image multiplied by the square root of the ratio of the solid angle of the relic to that of the synthesized beam. Since these two sources of errors are unrelated, they are added in quadrature to estimate the final error on the flux densities of the relics that are quoted in Table \ref{table:flux}.
    
The left and the right panels in Fig.~\ref{fig:spec_rel} show the spectra over the frequency range 80--1400 MHz, for the east and the west relics, respectively. The spectral index of the east relic is $-1.37\pm0.08$ while that of the west relic is $-1.17\pm0.06$. A least-squares fitting method was used to estimate the spectral index values from the flux density values given in Table~\ref{table:flux}. The reduced-$\chi^2$ values for the fits on the east and west relics were 1.7 and 3.4, respectively.

\section{Discussion}
Abell 3376 shows two arc-like radio relics on either side of the X-ray emission from the cluster. The arcs are oriented roughly perpendicularly to the merger axis of the cluster. Such a scenario is in agreement with the models that predict two outgoing shocks in the aftermath of the merger which are responsible for the radio emission seen from the arcs.

According to the DSA theory \citep{blandford87} the slope of the power law spectrum of the number density of electrons, $p$, as a function of energy (N(E) $\propto$ E$^p$), is related to the Mach number of the shock, $\mathcal{M}$, by

\begin{equation}
p = -2\frac{\mathcal{M}^2 + 1}{\mathcal{M}^2 - 1} - 1.
\end{equation}

\noindent This includes the effects of particle ageing due to inverse Compton and synchrotron radiation losses under the assumption of continuous injection \citep{sarazin99,giacintucci08}. The slope is related to the spectral index ($\alpha$) of the synchrotron radiation by

\begin{equation}
\alpha = \frac{p+1}{2}.
\end{equation} 

	The spectral index is related to the flux density, $S$, by the standard relation, $S\propto \nu^{\alpha}$, where $\nu$ is frequency. 
For the east relic, the spectral index is estimated to be $-1.37\pm0.08$ which gives the Mach number  $2.53\pm0.23$. For the west relic, the spectral index is estimated to be $-1.17\pm0.06$, which gives the Mach number $3.57\pm0.58$.

	$Suzaku$ X-ray observations by \citet{akamatsu12} estimate the Mach number of the East and West relics to be $2.91\pm0.91$. This number agrees with the value obtained from simulations \citep{machado13} as well. The Mach numbers estimated here for both the relics are consistent with the X-ray estimates.

	The spectra of the relics show no breaks in the frequency range 80 to 1400 MHz. The shock acceleration is expected to produce an initial injection index for the synchrotorn radiation close to -0.8. Radiation losses induce a break such that the spectral index after the break steepens by 0.5 to -1.3 (See for example, the relic radio galaxies in \citet{slee01}). The average spectral index of the two relics is $\sim$ -1.3. This is consistent with the break frequency being close to or lower than the lower end of the spectrum ($\sim$ 80 MHz) shown in Fig.~\ref{fig:spec_rel}. 
The magnetic field in the cluster can be assumed to be $\sim 1\mu$G \citep{govoni13}. We can estimate the spectral ages of the relics as,

\begin{equation}
t = 1060\frac{B^{0.5}}{B^2+B^2_{\rm IC}}[(1+z)\nu]^{-0.5},
\end{equation}
\noindent where $B$ is the magnetic field in the source, $B_{IC}$ is the effective Inverse Compton magnetic field, $z$ is the redshift of the source and $\nu$ is the break frequency \citep{slee01}. Based on these values, the spectral ages of the relics is $\sim$ 0.4 Gyr. 

We have made the assumption here that the break frequency is 80 MHz. In reality, this break could be lower than this which would imply that the spectral age is greater than 0.4 Gyr. This estimated spectral age of the relics is, nevertheless, consistent with the age of the cluster ($\sim$ 0.5 Gyr) as obtained from simulations discussed later in this section.
	
	\begin{figure*}
		\centering
		\includegraphics[width=0.5\textwidth,angle=270]{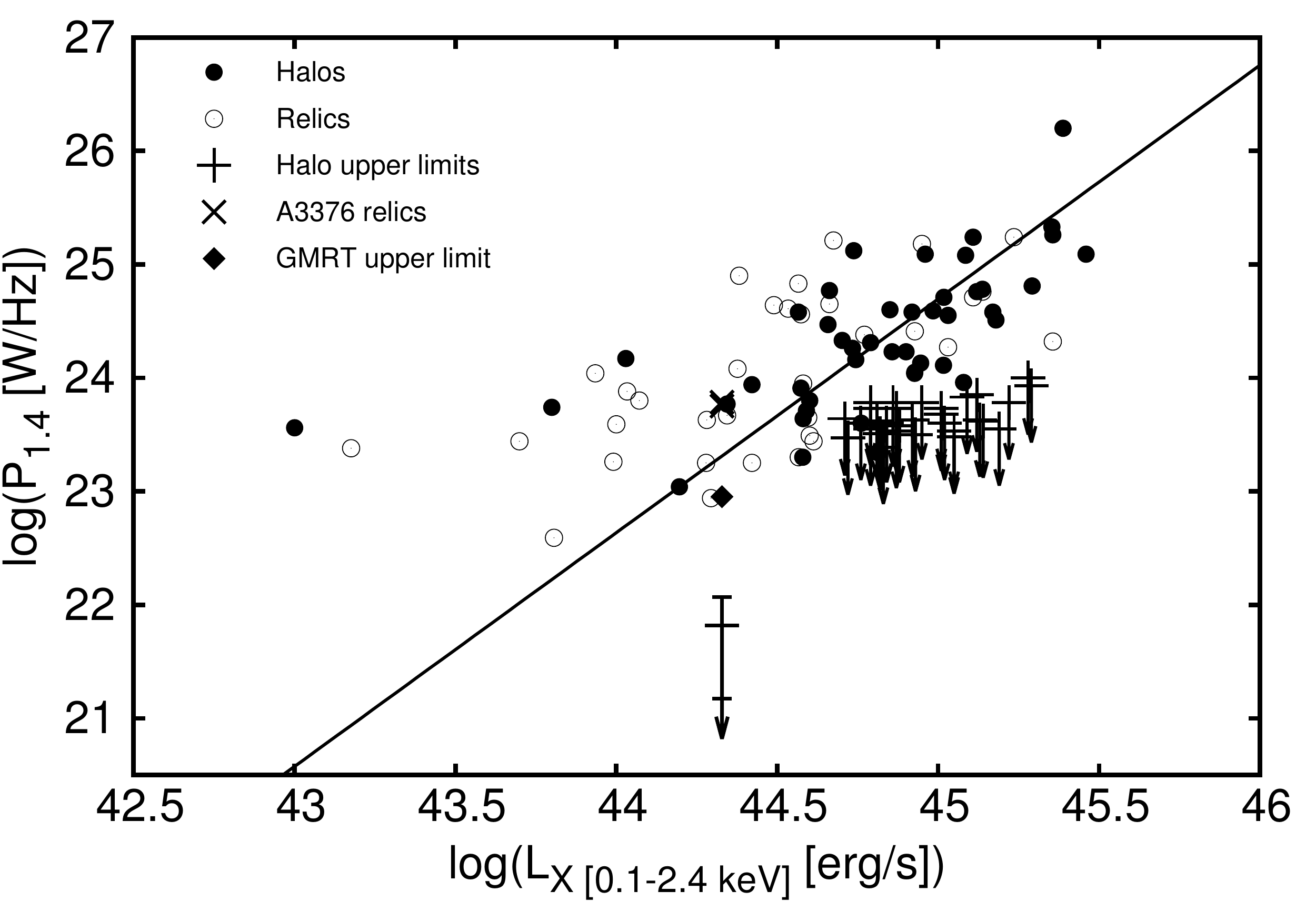}
		\caption{Plot showing the $L_X-P_{1.4}$ correlation for halos and relics. The filled circles represent all known radio halos. The empty circles represent radio relics \citep{feretti12}. The solid line represents the best fit for the $L_X-P_{1.4}$ relation for halos ($\textrm{log}(P_{1.4})-24.5 = 0.195 + 2.06[\textrm{log}(L_X) - 45]$, \citet{brunetti09}). The arrows represent upper limits on halo emission \citep{venturi08,kale13}. The cross represents the A3376 relics. The diamond symbol shows the upper limit to halo emission in A3376 as estimated by GMRT while the arrow below it represents the new upper limit as estimated by the MWA.}
		\label{fig:lxlr}
	\end{figure*}
	
	The cluster merger responsible for the generation of the observed relics also generates turbulence in the ICM which is believed to be able to re-accelerate electrons in the region between the relics \citep{petrosian01,brunetti01}. The region between the relics is then expected to host a radio halo. No such radio halo is detected at the MWA frequencies. However, we have been able to put an upper limit to the flux density from a possible halo in the central region of A3376. Table \ref{table:flux} lists the mean flux densities observed in the central regions of A3376 in the MWA images. The mean value of the spectral index of halos is $-1.34\pm0.28$ (\citet{feretti12}). An upper limit to the radio power of a halo at 1.4 GHz can be estimated using the upper limit at 154 MHz (Table \ref{table:flux}) and extrapolating it to 1400 MHz using the mean value of the spectral index quoted here. This power corresponds to 6 $\times$ 10$^{21}$ W Hz$^{-1}$. 
	
The upper limit is plotted in Fig~\ref{fig:lxlr} where the radio powers at 1.4 GHz and the corresponding X-ray luminosities of known halos and relics are also plotted \citep{brunetti09}. The solid line represents the best fit to the known halos. The upper limit on the radio power of the A3376 halo is lower by a factor of $\sim$35 compared to that expected from the best fit line.

       The upper limit on the radio power of the halo corresponds to a halo size of $\sim$ 160 kpc ($\sim 3'$) at 154 MHz. This corresponds to the beam size at 154 MHz. If the halo is more extended than this, even with higher total radio power, it will be undetected because of limits on surface brightness sensitivity. For a canonical halo as extended as 500 kpc, the upper limit on the radio power would still be lower by a factor of $\sim$ 4 as compared to that expected from the best-fit value.
 	
	In order to explain why no radio halo is detected in this cluster we first need to discuss some of the time scales involved in the merger process. The separation between the two relics of A3376 is $\sim 2$ Mpc. Simulations by \citet{machado13} estimate that the shock velocity, $v_{\rm s}$, is $\sim$ 2000 km s$^{-1}$ with an estimated dynamical timescale, $t_{\rm dyn}\sim$ 0.49 Gyr. This agrees with their simulations of the gas morphology which puts the cluster at approximately 0.5 Gyr after central passage. 
	
	MHD turbulence generated in cluster mergers has a decay time associated with it which is given by \citet{cassano05},
	
	\begin{equation}
		\tau_{\rm kk}(\text{Gyr}) \sim \left(\frac{v_{\rm i}}{2000 \text{ km s}^{-1}}\right)^{-1} \left(\frac{L_{\rm inj}}{1\text{ Mpc}}\right) \left(\frac{\eta_{\rm t}}{0.25}\right)^{-1},
	\end{equation}

\noindent where $v_{\rm i}$ is the relative velocity of the merging clusters at the time of impact and $\eta_{\rm t}$ is defined as the fraction of energy in the form of turbulence that is converted into magnetosonic (MS) waves. 
This means that the value of $\eta_{\rm t}$ defines the efficiency with which the electron population is re-accelerated. The value of $\eta_{\rm t}$ can be approximated to 0.25 by using the constraint that the slope of the radio spectrum of the halos is in the range, $\alpha = 1.1 - 1.5$ \citep{kempner01}. The value of $v_i$ is estimated from simulations \citep{machado13} to be $\sim 1500$ km s$^{-1}$. The typical injection length scale ($L_{\rm inj}$) of turbulence in galaxy clusters is $200-300$ kpc \citep{brulaz11}. With these values in the above equation, the decay time of turbulence for a cluster like 
A3376 is estimated to be $0.25-0.4$ Gyr. Considering that it is now $\sim$ 0.5 Gyr since core passage of the cluster, most of the turbulence that would have been generated at the time of merger
would have dissipated. Furthermore, the half-life of these radio emitting electrons is short, 
$\sim 0.3$ Gyr \citep{slee01}. Any radio halo that was generated during the cluster merger is likely dissipated.

It has been postulated that in galaxy clusters there should be a long-lived component of cluster-wide diffuse radio emission that is generated by relativistic electrons which are continuously injected into the IGM by relativistic proton - thermal proton collisions \citep{brulaz11}. The radio power of diffuse emission expected from these secondary particles is expected to be at a level that is about an order of magnitude lower than that of the radio powers in halos that are detected. The upper limit on the radio power of a halo in Abell 3376 is nearing such levels. While the current limit is close to such a level, further lowering of these limits either in this cluster or in other clusters will have implications for the secondary models of halo production.

\section{Summary}
	In this paper, we present low frequency radio observations of Abell 3376 using the Murchison Widefield Array at 88, 118, 154, 188 and 215 MHz. The observations were carried out as part of the GaLactic and Extragalactic All-sky MWA (GLEAM) survey. The observations were analysed using Common Astronomy Software Applications (CASA)
tools as well as specialised software packages written to process the GLEAM data. Images were produced at all frequencies. Towards the east and west peripheries of the cluster bright arcs of radio 
emission (relics) were detected at all frequencies. These arcs, with a linear extent of $\sim$ 1 Mpc each, are a result of outgoing shocks produced by the merger event and are separated by $\sim$ 1 Mpc from 
the cluster centre. Spectra for both the relics were estimated using the flux densities estimated from the MWA images along with those from previous observations of GMRT at 150 and 325 MHz and of VLA at 1400 MHz. The spectral indices of the east and west relics over the frequency range 80-1400 MHz are $-1.37\pm0.08$ and  $-1.17\pm0.06$ respectively. The Mach numbers of the shocks estimated from these indices are consistent with those estimated from ${\it Suzaku}$ X-ray observations. Assuming that the break frequency due to radiation losses corresponds to the lower end of the spectrum, the age of the relics can be estimated to be $\sim$ 0.37 Gyr. This age is consistent with the age of the cluster estimated from simulations to be $\sim$ 0.5 Gyr.
   
   No diffuse radio emission was detected in the region between the relics. The upper limit on the radio power of any possible halo of 160 kpc size is $\sim$ 6$\times$ 10$^{21}$ W Hz$^{-1}$ at 1.4 GHz. This is a factor of 35 lower than that expected from the correlation between radio powers and X-ray luminosities of cluster halos. It is very likely that the radio emission from the halo has substantially decayed over its age ($\sim$ 0.37 Gyr) considering that this time scale is comparable to the life time of the radiating electrons at the MWA observing frequencies. The upper limit on the radio power of a halo in Abell 3376 is a factor of 10 lower than most upper limits available so far and is nearing levels at which radio emission from secondary electrons produced in relativistic proton - thermal proton collisions is expected. 
Given the uncertainties in the hadronic model \citep{brulaz11} it is not possible to discuss the validity of the model based on the observations of the cluster alone. However, observations of a large number of clusters with MWA and LOFAR at the current limits of detection or using the SKA at improved sensitivities will have implications on the hadronic model.
   
\section*{Acknowledgement}
This scientific work makes use of the Murchison Radio-astronomy Observatory, operated by CSIRO. We acknowledge the Wajarri Yamatji people as the traditional owners of the Observatory site. Support for the MWA comes from the U.S. National Science Foundation (grants AST-0457585, PHY-0835713, CAREER-0847753, and AST-0908884), the Australian Research Council (LIEF grants LE0775621 and LE0882938), the U.S. Air Force Office of Scientific Research (grant FA9550-0510247), and the Centre for All-sky Astrophysics (an Australian Research Council Centre of Excellence funded by grant CE110001020). Support is also provided by the Smithsonian Astrophysical Observatory, the MIT School of Science, the Raman Research Institute, the Australian National University, and the Victoria University of Wellington (via grant MED-E1799 from the New Zealand Ministry of Economic Development and an IBM Shared University Research Grant). The Australian Federal government provides additional support via the Commonwealth Scientific and Industrial Research Organisation (CSIRO), National Collaborative Research Infrastructure Strategy, Education Investment Fund, and the Australia India Strategic Research Fund, and Astronomy Australia Limited, under contract to Curtin University. We acknowledge the iVEC Petabyte Data Store, the Initiative in Innovative Computing and the CUDA Center for Excellence sponsored by NVIDIA at Harvard University, and the International Centre for Radio Astronomy Research (ICRAR), a Joint Venture of Curtin University and The University of Western Australia, funded by the Western Australian State government.


\bibliographystyle{mn2e}

\label{lastpage}
\end{document}